\documentclass[twocolumn, showkeys, superscriptaddress, amsmath, amssymb, aps, pra, floatfix, reprint]{revtex4-2}

\usepackage{graphicx}
\usepackage{dcolumn}
\usepackage{bm}
\usepackage{hyperref}
\usepackage{xcolor,tikz, circuitikz}

\usetikzlibrary{positioning, shapes.geometric, shapes.symbols, decorations.pathmorphing}

\hypersetup{
colorlinks,
linkcolor={blue},
citecolor={blue},
urlcolor={blue!80!black}
}

\usepackage{amsthm}
\usepackage{multirow}

\begin{document}

\title{Nonuniform Time-bins for Making Raw QRNG Outputs Uniform}

\author{Nirupam Basak}
\email{nirupambasak2020@iitkalumni.org}
 \affiliation{Indian Statistical Institute, Kolkata 700108, India.}
\author{Goutam Paul}%
 \email{goutam.paul@isical.ac.in}
 \affiliation{Indian Statistical Institute, Kolkata 700108, India.}

\begin{abstract}
Random numbers play a vital role in scientific simulations and cryptography. Both randomness and security are critical requirements for random number generators (RNGs). However, classical RNGs often rely on mathematical assumptions that can compromise security, while quantum RNGs (QRNGs) face challenges, including non-uniform output due to the lack of specialized single-photon sources. In this work, we propose a QRNG model that generates uniformly distributed random numbers using a standard coherent photon source, without requiring heavy attenuation or post-processing. Although the raw data is theoretically free of bias, post-processing, if any, may be required to suppress any classical noise and device imperfections. The proposed method analytically equalizes the output probabilities under the assumed photon-arrival-time model, potentially reducing the need for heavy attenuation and post-processing to correct binning bias. We evaluate the method using real photon-detection timestamps and compare its statistical performance with conventional uniform time binning using the NIST, Dieharder, and ENT test suites. The results demonstrate improved statistical performance for the proposed binning scheme under the tested conditions.
\end{abstract}

\keywords{Quantum random number generator, Coherent photon source, Uniform randomness}
\maketitle

\section{Introduction}

Nowadays, generating true random numbers is crucial for scientific simulations~\cite{carlson2015quantum,bell1964on,nicholas1949the}, cryptographic applications~\cite{shannon1949communication,charles1992quantum} and digital games~\cite{konstantinou2004electronic,yamanaka2017enumeration}. However, classical pseudo-random number generators (PRNGs) are deterministic by nature and therefore cannot guarantee security in the domains where they are used. In contrast, a true random number generator (TRNG) generates random numbers from physical entropy sources~\cite{schindler2009random,stipčević2014true, ma2016quantum,mannalatha2023comprehensive}. A quantum random number generator (QRNG) is one such TRNG that uses the properties of quantum particles and the principles of quantum mechanics~\cite{ma2016quantum,mannalatha2023comprehensive} to produce random numbers. Quantum theory asserts that these numbers are inherently random, making them unpredictable.

Various quantum random number generator (QRNG) models have been proposed based on distinct quantum phenomena, including quantum superposition~\cite{jennewein2000fast,stefanov2000optical,kwon2009quantum,xu2018pixel,grafe2014chip}, optical pulses~\cite{furst2010high,ren2011quantum,applegate2015efficient,sanguinetti2014quantum,qi2010highspeed,xu2012ultrafast}, single-photon detection~\cite{wang2006scheme,nie2014practical,dynes2008high,wayne2009photon,wahl2011ultrafast,zhang2021simple}, vacuum fluctuations~\cite{shen2010practical,gabriel2010generator,xiao1987precision,haw2015maximization,guo2019parallel}, spontaneous emission~\cite{marosits2020amplified, guo202140gbps} and quantum tunneling~\cite{bernardo2017extracting, aungskunsiri2021random,liu2023highrandomness}. Additionally, some designs leverage different types of quantum noise, such as shot noise, phase noise, and spin noise~\cite{qi2010highspeed, guo2010truly, shen2010practical, xu2012ultrafast, nie2015generation, yang20165.4gbps, huang2020phase}. Among these, photon-based optical QRNGs are particularly promising due to their relative ease of implementation~\cite{ma2016quantum,mannalatha2023comprehensive}. This advantage originates from the extensive development of optical technologies and the widespread availability of optical components.

In recent years, several photon-based QRNGs have been proposed, utilizing methods such as beam splitters~\cite{jennewein2000fast,stefanov2000optical,kwon2009quantum,xu2018pixel,grafe2014chip,rarity1994quantum}, photon counting~\cite{tisa2015high,ren2011quantum,applegate2015efficient,balygin2018quantum,sanguinetti2014quantum}, and photon arrival time~\cite{furst2010high,nie2014practical,khanmohammadi2015monolithic,wayne2010lowbias,dynes2008high,wayne2009photon,wahl2011ultrafast,stipčević2007quantum,zhang2021simple,ma2005random,yan2015high}. These approaches typically require precise preparation of single photons and highly efficient detection systems. However, the technical difficulties associated with developing reliable single-photon sources and single-photon-resolving detectors present significant obstacles to the practical implementation of such QRNGs~\cite{imamoglu1994turnstile, kim1999single}. To address these issues, researchers have proposed photon arrival-time-based QRNGs employing attenuated coherent laser sources~\cite{nie2014practical,wahl2011ultrafast,zhang2021simple,stipčević2007quantum,ma2005random}. While this method circumvents the need for true single-photon sources and photon-number-resolving detectors, the generated random numbers inherently follow a non-uniform distribution~\cite{nie2014practical}.

Since the underlying distribution of the generated random numbers is typically non-uniform, post-processing is required to extract the inherent entropy. This post-processing involves systematically compressing the data to extract the entropy~\cite{dichtl2007bad,balakrishnan2007relationship}. However, such compression reduces the random number generation rate, as it decreases the amount of usable data within a given time period. Moreover, because these post-processing techniques are classical and deterministic, they may introduce potential security vulnerabilities~\cite{kuznetsov2019side,park2019scrqrng,chowdhury2021physical}. Consequently, post-processing should ideally be avoided or minimized. To achieve this, it is crucial to generate raw data with the highest possible entropy at the source. Nguyen et al.~\cite{nguyen2018programmable} proposed a postprocessing-free QRNG using customized pulse shapes. However, this requires heavy attenuation of the photon pulses. The use of multiple attenuators to produce weak coherent pulses increases the cost of the QRNG. Moreover, as such attenuation increases the probability of having no photon in a time interval, it reduces the generation rate of the random numbers~\cite{paul2025evaluating}.

In this article, we propose a random number generation method based on coherent light and a photon detector, capable of producing raw data with an inherently uniform distribution. Random numbers are generated from the photon arrival times. While the core concept resembles existing photon arrival-time-based QRNGs, the proposed model theoretically generates uniform random numbers without requiring any post-processing. However, in realistic experimental settings, quantum randomness is unavoidably contaminated by classical noise, e.g., device imperfections~\cite{xu2012ultrafast, liu2025postprocessing}. Therefore, an appropriate post-processing procedure may be required to remove any correlations between the generated random bits and the classical noise. Another advantage of our approach is that it can be implemented without heavily attenuating the coherent pulses, resulting in a higher random number generation rate compared to existing models. Additionally, it does not require a highly efficient detector; however, a lower detector efficiency would reduce the generation rate. We further validate the proposed model using practical timestamp data and demonstrate that the resulting random numbers achieve superior statistical performance compared with the conventional uniform bin-based method, as evaluated using the NIST~\cite{bassham2010sp}, Dieharder~\cite{brown2014dieharder}, and ENT~\cite{walker2008ent} statistical test suites.%

Unlike existing approaches that address output bias through optical pulse shaping~\cite{wayne2010lowbias,nguyen2018programmable} or bounded-bias techniques~\cite{wahl2011ultrafast}, our approach compensates for the nonuniform photon-arrival-time distribution through analytically designed nonuniform time bins. We derive closed-form bin boundaries that equalize the probability of the first registered detection across all bins, conditioned on at least one detection within a reference cycle. Thus, our contribution lies in the time-to-number mapping rather than optical pulse engineering. A detailed comparison is presented in Appendix~\ref{app:comparison_qrng}.

The remainder of the article is organized as follows. Section~\ref{sec:existing_qrng} briefly discusses existing QRNG models. Section~\ref{sec:proposal} presents the proposed nonuniform time-binning scheme for equalizing detection probabilities across bins. Section~\ref{sec:numerics} presents the experimental results and statistical tests of the uniformity of the raw outputs. Finally, Section~\ref{sec:conclusion} discusses the findings and concludes the article.

\section{\label{sec:existing_qrng}Existing photon-arrival-time-based QRNG}

Photon-arrival-time-based quantum random number generators (QRNGs) utilize weak coherent pulses as photon sources, where the detection of a photon by a single-photon detector (SPD) is timestamped to derive random numbers. To ensure that the probability of multiple photons arriving within a given time interval remains negligible, the coherent pulses must be heavily attenuated~\cite{wayne2010lowbias,dynes2008high,wayne2009photon,wahl2011ultrafast,stipčević2007quantum,zhang2021simple,ma2005random,yan2015high}. However, a fundamental limitation of this approach arises from the photon statistics of coherent light, which follow a Poisson distribution. As a result, the generated photon arrival times are not uniformly distributed, leading to non-uniform randomness in the raw output. This necessitates additional post-processing to extract near-uniform random bits, which may reduce throughput and increase vulnerability to side-channel attacks~\cite{kuznetsov2019side,park2019scrqrng,chowdhury2021physical}.

\subsection{Existing Ideal Model}

Three principal models have been proposed over time for coherent-pulse-based quantum random number generators (QRNGs), each utilizing a distinct mapping from photon arrival times to random output. These schemes harness the stochastic behavior of photon detection events to generate random bit sequences. These three models are as follows.

\begin{enumerate}
\item[(I)]\emph{Inter-arrival Time Comparison Model:} This method compares consecutive photon inter-arrival times, denoted as $\Delta t_{2i-1}$ and $\Delta t_{2i}$, to derive the $i$-th bit of the random number. Specifically, the output is assigned as $0$ if $\Delta t_{2i-1} < \Delta t_{2i}$ and $1$ otherwise. While this approach provides statistically sound randomness, it has a low bit generation rate, producing one bit per two detection events~\cite{dynes2008high, schranz2024stochastic}.
\item[(II)]\emph{Fixed Sleep Time Model:} In this model, the detection timeline is partitioned into consecutive time bins of equal duration $\Delta t$, indexed as $0, 1, 2, \dots$. The index of the bin in which a photon is detected is used to generate the corresponding output in binary form~\cite{wayne2009photon,wayne2010lowbias,wahl2011ultrafast,stipčević2007quantum}. After each detection event, the detector enters a predefined inactive period, known as the sleep time $t_s$, after which bin indexing restarts.
\item[(III)]\emph{External Clock Referencing Model:} This widely adopted~\cite{ma2005random,nie2014practical,wayne2010lowbias} method employs an external clock cycle of fixed duration $T$, divided into $N$ equal sub-intervals, called time bins, indexed from $0$ to $N-1$. In each cycle, the index of the time bin in which the first photon is detected is used as the random output after converting it into a binary number.
\end{enumerate}

Under ideal conditions, such as perfect photon detection, exactly one photon per cycle, etc., these models produce output distributions that are close to uniform. However, in practical implementations, imperfections such as detector inefficiencies and the presence of multi-photon events introduce statistical biases~\cite{nie2014practical}.

From now on, the discussion will be based on the QRNG using an external clock. However, this discussion is also applicable to the QRNG based on fixed sleep time.

\subsection{Existing Model in Practical Scenario}

Photon number distribution of coherent pulse follows Poisson distribution~\cite{titilaer1966density}. The probability of having $n$ photons within time $T$ is given by
\begin{equation}
\label{E:Psn}
P(n,T):=\frac{e^{-\mu T}(\mu T)^n}{n!},
\end{equation}
where $\mu$ is the mean photon number in unit time. A series of attenuators reduces the power of the pulse. This reduces the value of $\mu$. As $P(n,T)>0$ for all $n\geq0$ when $\mu>0$, for any positive $\mu$, the probability of having multiple photons in time $T$ cannot be $0$. However, setting $T$ less than the \emph{dead-time}, which is the minimum time gap between two consecutive detections of the detector, one can enforce this probability to be 0 for very small $\mu$. Thus, one requires heavy attenuation of the coherent photon pulses. Such attenuation results in a high cost.

Also, the detectors are not perfect. The detection efficiency also has a crucial role in the final distribution. Considering these two faults, we can write the probability that the first photon would be detected in the $i$-th bin as~\cite{nie2014practical}
\begin{align}
\label{E:1st_det}
P(i;\eta;1st)%
&=(e^\frac{\mu T\eta}{N}-1)e^{-i\frac{\mu T\eta}{N}}.
\end{align}
This, along with the probability of no detection in time $T$, $P_\eta(0,T)=e^{-\mu T\eta}$, gives the distribution of photon detection. Here by
\begin{equation}
\label{E:psn_eta}
P_\eta(n,T):=\frac{e^{-\mu T\eta}(\mu T\eta)^n}{n!}
\end{equation}
we denote the probability of detecting $n$ photons with detection efficiency $\eta$ within a time interval $T$. Thus, we can write the probability mass function (p.m.f.) for the generated random number as~\cite{paul2025evaluating}
\begin{equation}
\label{E:p.m.f.}
f(i)=\frac{e^\frac{\mu T\eta}{N}-1}{1-e^{-\mu T\eta}}e^{-i\frac{\mu T\eta}{N}},\;i=1,2,\dots,N.
\end{equation}
Note that this being a decreasing function, such a QRNG produces smaller numbers with high probabilities and bigger numbers with low probabilities. Therefore, the produced random numbers are not uniform and require post-processing.

In addition, smaller $\mu$ reduces the random number generation rate by increasing the probability of no photon detection $P_\eta(0,T)$~\cite{paul2025evaluating}. To overcome all of these drawbacks, we propose a new QRNG model.

\section{\label{sec:proposal}New Proposal for QRNG}

The p.m.f.~\eqref{E:p.m.f.} for the raw number generated by the existing QRNGs being a decreasing function indicates that more photons have been detected in the bins with lower indices compared to higher indices. One possible pathway to make this uniform is by varying the length of bins, smaller for low indices and higher for high indices. Suppose $l_i,\ i=0,1,\dots,N-1$ be the length of $i$-th time bin. Let us define
\begin{equation}
\label{E:len_sum}
\begin{aligned}L_{i+1}&:=\sum\limits_{j=0}^il_j\text{ for }i=0,1,\dots,N-1\;\;L_0:=0.
\end{aligned}
\end{equation}

To produce a uniform random distribution, we have to find $l_i$ such that the probability of detecting the first photon in the $i$-th bin is constant $C$ for all $i$. Let us assume that $n$ photons are present in time $T$. Then, considering the bin length as in Eq.~\eqref{E:len_sum}, the probability that the first photon would be detected in the $i$-th bin is given by
\begin{widetext}
\allowdisplaybreaks
\begin{align*}
P^{(l)}(i;1st|n)
&:=\frac{\sum\limits_{k=1}^nP_\eta(0,L_i)P_\eta(k,l_i)P_\eta(n-k,T-L_{i+1})}{P_\eta(n,T)}\\
&=\frac{\sum\limits_{k=1}^ne^{-\mu L_i\eta}\frac{e^{-\mu l_i\eta}(\mu l_i\eta)^k}{k!}\frac{e^{-\mu(T-L_{i+1})\eta}(\mu(T-L_{i+1})\eta)^{(n-k)}}{(n-k)!}}{e^{-\mu T\eta}\frac{(\mu T\eta)^n}{n!}}\\
&=\sum_{k=1}^n\frac{n!}{k!(n-k)!}\left(\frac{l_i}{T}\right)^k\left(1-\frac{L_{i+1}}{T}\right)^{n-k}\\
&=-\left(1-\frac{L_{i+1}}{T}\right)^n+\sum_{k=0}^n\frac{n!}{k!(n-k)!}\left(\frac{l_i}{T}\right)^k\left(1-\frac{L_{i+1}}{T}\right)^{n-k}\\
&=-\left(1-\frac{L_{i+1}}{T}\right)^n+\left(\frac{l_i}{T}+1-\frac{L_{i+1}}{T}\right)^n=\left(1-\frac{L_i}{T}\right)^n-\left(1-\frac{L_{i+1}}{T}\right)^n.
\end{align*}
\end{widetext}
Therefore, the probability of detecting the first photon in the $i$-th bin would be
\begin{widetext}
\allowdisplaybreaks
\begin{align*}
C=P^{(l)}(i;1st)
&:=\sum_{n\geq0}P^{(l)}(i;1st|n)P_\eta(n,T)\\
&=\sum_{n\geq0}\left[\left(1-\frac{L_i}{T}\right)^n-\left(1-\frac{L_{i+1}}{T}\right)^n\right]e^{-\mu T\eta}\frac{(\mu T\eta)^{n}}{n!}\\
&=e^{-\mu T\eta}\sum_{n\geq0}\frac{1}{n!}\left[\left\{\mu T\eta\left(1-\frac{L_i}{T}\right)\right\}^n-\left\{\mu T\eta\left(1-\frac{L_{i+1}}{T}\right)\right\}^n\right]\\
&=e^{-\mu T\eta}\left(e^{\mu T\eta\left(1-\frac{L_i}{T}\right)}-e^{\mu T\eta\left(1-\frac{L_{i+1}}{T}\right)}\right)\\
&=e^{-\mu L_i\eta}-e^{-\mu L_{i+1}\eta}.
\end{align*}
\end{widetext}
From this, we get the recurrence relation for $L_i$ as
\begin{equation}
\label{E:rec_rel}
\begin{aligned}
&L_{i+1}=-\frac{\ln\left(e^{-\mu L_i\eta}-C\right)}{\mu\eta},\ i=0,1,\dots,N-1\\
&L_0=0.
\end{aligned}
\end{equation}
We can prove by mathematical induction that the solution to this recurrence relation is given by
\begin{equation}
\label{E:rec_sol}
L_i=-\frac{\ln\left(1-iC\right)}{\mu\eta}\text{ for }i=0,1,\dots,N.
\end{equation}
From this, the  length of the $i$-th bin can be written as
\begin{equation}
\label{E:len_bin}
l_i=L_{i+1}-L_i=\frac{1}{\mu\eta}\ln\left(\frac{1-iC}{1-(i+1)C}\right).
\end{equation}

Now, by definition~\eqref{E:len_sum}, we have $L_N=T$. Combining this with~\eqref{E:rec_sol} we get
\begin{equation}
C=\frac{1-e^{-\mu T\eta}}{N}.
\label{E:const_val}
\end{equation}
Note that $C\neq\frac{1}{N}$. The sum of this probability over $N$ time-bins gives $1-e^{-\mu T\eta}$. The remaining probability, $e^{-\mu T\eta}$ is the probability of no photon detection $P_\eta(0,T)$ within time $T$. Thus, the p.m.f. for the generated random number becomes
\begin{equation}
\label{E:pmf_uni}
f_U(i)=\frac{C}{1-P_\eta(0,T)}=\frac{1}{N},\ i=0,1,\dots,N-1
\end{equation}

Equation (10) establishes a uniform distribution of detection times across the bins, conditioned on at least one registered detection within the reference cycle. The result assumes that the effective detection rate, $\mu\eta$, is constant and that the adopted first-arrival-time model accurately describes the detection process. It does not, by itself, establish independence between successive outputs or exclude correlations arising from detector imperfections and fluctuations in the source intensity. Consequently, the theoretical uniformity result should be distinguished from the randomness and entropy of the output generated by a practical implementation.

Observe that $b = \log_2N$ bits of random number, given by the index $i$ of the time-bin in which the photon has been detected, can be generated from each photon detection. The probability that $b$ bits of random numbers would be generated in a cycle is given by the probability of detecting a photon within that cycle, $P(success)=NC$. Therefore, the rate for generating random numbers that can be achieved with this model is given by
\begin{equation}
\label{E:rate}
r=\frac{NC}{T}b=\frac{1-e^{-\mu T\eta}}{T}\log_2N\text{ bits per unit time}.
\end{equation}
Note that this is the same rate as the existing uniform bin-length-based approaches~\cite{paul2025evaluating}. Moreover, in the proposed model, the mean photon number $\mu$ does not influence the distribution of the generated numbers. As a result, the model theoretically eliminates the compulsory attenuation of coherent pulses. In practice, it reduces attenuation, thereby increasing the random number generation rate. However, classical noise and device imperfections like deviation of the mean photon number $\mu$ and detection efficiency $\eta$, may add some bias. Thus, appropriate post-processing may be required in practice to extract the minimum entropy. Suppose the minimum entropy is $H_\mathrm{min}$ bits per bit. Then the effective rate would be bounded above by $rH_\mathrm{min}$, i.e.,
\begin{equation}
\label{E:rate_eff}
r_\mathrm{eff}<H_\mathrm{min}\frac{1-e^{-\mu T\eta}}{T}\log_2N\text{ bits per unit time}.
\end{equation}

\begin{figure*}
\centering
\includegraphics[width=0.9\textwidth]{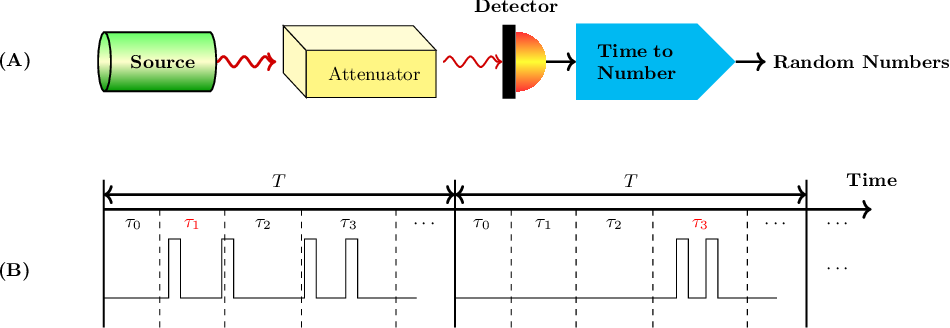}
\caption{(A) Schematic diagram of the QRNG Model, which is same as the existing model but with less attenuation. Photons coming from a coherent photon source are detected by a detector. Random numbers are generated from the detection time. (B) Time is divided into fixed-length cycles of duration $T$, each split into $N$ non-uniform bins, $\tau_0,\tau_1,\dots$. If the registered time of the first photon detection in a cycle falls in $\tau_i$, the corresponding random number would be $i-1$. In the first cycle, the first detection occurs in $\tau_1$, giving a random number of $1$. Similarly, the number given by the second cycle would be $3$, since the first detection occurred at $\tau_3$.}
\label{fig:model}
\end{figure*}

Also, the existing detectors may not be used with a completely unattenuated photon source. Therefore, we must require some amount of attenuation. However, with this non-uniform binning, we do not need to ensure the existence of at most one photon in each cycle, which was a compulsory requirement for the existing uniform bin-based method. In the proposed scheme, only the first registered detection within each reference cycle is used to generate the output symbol. Subsequent detections within that cycle are not used. This mapping avoids requiring the presence of at most one photon within a cycle; however, it does not eliminate the physical effects of multiple-photon events on the detector. The operating regime must therefore account for the detector's maximum count rate, dead time, saturation characteristics, and timing resolution.

We can summarize the proposed QRNG model as follows.
\begin{itemize}
\item\emph{Source: }Coherent pulses coming from a photon source.
\item\emph{Detector: }Photo-detector detects photons and registers the detection time.
\item\emph{Time-to-number mapping: }Time is divided into cycles of fixed length $T$, which works as reference time. Each of these cycles would be divided into $N$ bins such that the length of the $i$-th bin would be $l_i$ given by~\eqref{E:len_bin}. For each photon detection, the time would be recorded, and the detector would enter sleep mode until the end of the reference cycle. If the registered time falls in the $i$-th bin, the corresponding random number would be the $\log_2N$ bit binary representation of $i$.
\end{itemize}
A schematic diagram of the model is given in Fig.~\ref{fig:model}.

\section{\label{sec:numerics}Experimental Results and Statistical Tests}

We used two sets of raw photon-detection timestamps collected from the photon-arrival-time QRNG at DYSL-QT, Pune. For each dataset, we generated random-number sequences using both the existing time-binning method and the proposed nonuniform time-binning method. This enabled a comparison of the two approaches using the same experimentally acquired timestamp data, thereby isolating the effect of the binning strategy on the resulting output. The physical parameters are given in Table~\ref{tab:params}.

\begin{table}[htpb]
\centering
\caption{\label{tab:params}Physical parameters for data generation}
\begin{tabular}{|l|l|}
\hline
Number of detected photons&$17\times10^6$ per second\\
Detection efficiency&$54\%$\\
External cycle length&40 nano-second\\
Number of bins&1024\\
\hline
\hline
Generated data&10 bits per detection\\
\hline
\end{tabular}
\end{table}

The characteristics of the datasets, including their size, time span, generation speed of raw data, and entropy, are summarized in Table~\ref{tab:datasets}. The higher Shannon entropy and minimum entropy values observed in the data generated using the proposed non-uniform binning method, compared with those obtained using conventional uniform binning, indicate that non-uniform binning can enhance the randomness of the generated data.

\begin{table}[htpb]
\centering
\caption{\label{tab:datasets}Characteristics of the generated data from the real time-stamps}
\resizebox{\columnwidth}{!}{\begin{tabular}{|c|c|c|c|c|}
\hline
\multirow{3}{*}{\textbf{Characteristics}}&\multicolumn{2}{c|}{dataset 1}&\multicolumn{2}{c|}{dataset 2}\\
\cline{2-5}
&\textbf{Uniform}&\textbf{Non-uniform}&\textbf{Uniform}&\textbf{Non-uniform}\\
&\textbf{bins}&\textbf{bins}&\textbf{bins}&\textbf{bins}\\
\hline
\textbf{Data size (Gigabits)}&\multicolumn{2}{c|}{$\thicksim6.64$}&\multicolumn{2}{c|}{$\thicksim6.29$}\\
\textbf{Time span (Seconds)}&\multicolumn{2}{c|}{$\thicksim40$}&\multicolumn{2}{c|}{$\thicksim40$}\\
\textbf{Raw speed (Mbps)}&\multicolumn{2}{c|}{$\thicksim$ 165.80}&\multicolumn{2}{c|}{$\thicksim$ 157.15}\\
\hline
\textbf{Entropy (bits/Byte)}&7.999994&\textbf{7.999999}&7.999994&\textbf{7.999999}\\
\textbf{Min. entropy}&7.977&\textbf{7.996}&7.977&\textbf{7.995}\\
\hline
\end{tabular}}
\end{table}

The generated data were evaluated using multiple statistical test suites, including NIST, Dieharder, and ENT, to assess their randomness. NIST and Dieharder evaluate the data using a range of statistical tests, with results interpreted according to the corresponding $p$-values and test-specific criteria. The NIST report contains 188 individual test-result entries across 15 statistical test types, with evaluation based on both the uniformity of $p$-values and the proportion of passing sequences. The Dieharder report, in contrast, comprises 114 individual test-result entries across 31 statistical test types, each assigned a $p$-value and classified as PASSED, WEAK, or FAILED. ENT provides a complementary assessment by reporting statistical measures such as entropy, arithmetic mean, the Monte Carlo estimate of $\pi$, serial correlation, etc.

\begin{table}[htpb]
\centering
\caption{\label{tab:stat_test}Results indicating the percentage of test results that the tested data has passed (or shown weak for Dieharder), from different statistical tests. 
Note that the proposed non-uniform time-bin-based model outperforms the existing uniform time-bin-based model. 
Since no post-processing has been applied to these data, the results indicate that the raw data generated by the proposed model are more uniform than those from the existing uniform binning method.
}
\resizebox{\columnwidth}{!}{\begin{tabular}{|c|c|c|c|c|}
\hline
\multirow{2}{*}{\textbf{Testing}}&\multicolumn{2}{c|}{dataset 1}&\multicolumn{2}{c|}{dataset 2}\\
\cline{2-5}
\multirow{2}{*}{\textbf{Suites}}&\textbf{Uniform}&\textbf{Non-uniform}&\textbf{Uniform}&\textbf{Non-uniform}\\
&\textbf{bins}&\textbf{bins}&\textbf{bins}&\textbf{bins}\\
\hline
\textbf{NIST}&98\%&\textbf{100\%}&98\%&\textbf{100\%}\\
\textbf{Dieharder}&50\%(4\%)&\textbf{86\%(5\%)}&48\%(9\%)&\textbf{88\%(4\%)}\\
\hline
\end{tabular}}
\end{table}

The test results for NIST and Dieharder have been provided in Table~\ref{tab:stat_test}. Observe that the data generated using the proposed non-uniform time-bin-based model outperforms the existing uniform time-bin-based model. Although the results for the NIST suite are very close for both models, the result is very clear from the Dieharder suite, where the uniform bin-based data passed less than $50\%$ of tests; in contrast, the proposed non-uniform bin-based data passed $88\%$ of tests and failed only $8\%$. The results for the ENT suite have been presented in Table~\ref{tab:ent}. This shows the same trend as the other two suites.

\begin{table}[htpb]
\centering
\caption{\label{tab:ent}Results of tests from ENT suite. 
The values for the non-uniform time-bin-based data, which are very close to the ideal values for the corresponding tests, indicate that the data are nearly random. On the other hand, the values for the existing uniform time-bin-based model are far from the ideal value. These results validate the improved randomness in the data generated by the proposed model.
}
\resizebox{\columnwidth}{!}{\begin{tabular}{|c|c|c|c|c|c|}
\hline
\multirow{3}{*}{\textbf{Tests}}&\multirow{2}{*}{\textbf{Ideal}}&\multicolumn{2}{c|}{dataset 1}&\multicolumn{2}{c|}{dataset 2}\\
\cline{3-6}
&\multirow{2}{*}{\textbf{values}}&\textbf{Uniform}&\textbf{Non-uniform}&\textbf{Uniform}&\textbf{Non-uniform}\\
&&\textbf{bins}&\textbf{bins}&\textbf{bins}&\textbf{bins}\\
\hline
\textbf{Entropy}&8.00&7.999994&\textbf{7.999999}&7.999994&\textbf{7.999999}\\
\hline
\textbf{Arithmetic}&\multirow{2}{*}{127.50}&\multirow{2}{*}{127.42}&\multirow{2}{*}{\textbf{127.50}}&\multirow{2}{*}{127.42}&\multirow{2}{*}{\textbf{127.50}}\\
\textbf{Mean}&&&&&\\
\hline
\textbf{Monte Carlo}&\multirow{2}{*}{3.1416}&\multirow{2}{*}{3.1429}&\multirow{2}{*}{\textbf{3.1414}}&\multirow{2}{*}{3.1429}&\multirow{2}{*}{\textbf{3.1414}}\\
\textbf{Value of $\pi$}&&&&&\\
\hline
\textbf{Serial}&\multirow{2}{*}{0.00}&\multirow{2}{*}{0.000026}&\multirow{2}{*}{\textbf{0.000007}}&\multirow{2}{*}{\textbf{-0.000011}}&\multirow{2}{*}{-0.000036}\\
\textbf{Correlation}&&&&&\\
\hline
\end{tabular}}
\end{table}

The statistical tests indicate that, for the timestamp datasets examined, nonuniform time binning generally performs better than conventional uniform binning according to the reported test results. The improvement is particularly apparent in the Dieharder results, whereas the NIST results are close for both methods. The ENT statistics also show that several measured quantities for the nonuniformly binned data are closer to their ideal reference values. These findings demonstrate improved statistical behavior under the tested conditions.

\section{\label{sec:conclusion}Discussion}

We have proposed a QRNG model that uses nonuniform time-bin boundaries to equalize the output probabilities under the photon-arrival-time model. The construction provides a theoretical means of reducing the binning bias associated with conventional uniform time bins. However, it does not eliminate the detector's physical operating constraints, and accordingly, some post-processing may be required in a practical implementation.

We evaluated the proposed mapping using real photon-detection timestamps, and compared its statistical performance with that of conventional uniform binning using the NIST, Dieharder, and ENT test suites. The results show improved statistical performance for the proposed mapping under the tested conditions. Nevertheless, statistical tests alone cannot establish the security of the generated bits or quantify the entropy of the physical source. Practical deployment, therefore, requires appropriate characterization of detector behavior, calibration stability, residual correlations, and output entropy. The proposed method should be viewed as a binning strategy for improving the uniformity of raw QRNG outputs, with its practical throughput and security determined by the characteristics of the complete implementation.

\begin{acknowledgments}
We wish to acknowledge the support of the DYSL-QT, Pune, India, in sharing the timestamp data and the corresponding physical parameters used to test the proposed model.
\end{acknowledgments}

\section*{Data Availability Statement}The data are available from the corresponding author upon reasonable request, subject to permission from the data provider.

\bibliographystyle{unsrt}
\bibliography{ref}

\appendix

\section{Distinction of Present Work from the Existing QRNGs}
\label{app:comparison_qrng}
Several approaches have been proposed to reduce bias in photon-arrival-time-based QRNGs, including shaped optical pulses~\cite{wayne2010lowbias}, bounded-bias techniques~\cite{wahl2011ultrafast}, and programmable post-processing-free generation~\cite{nguyen2018programmable}. In contrast, our work addresses bias through analytically designed nonuniform time bins. The derived bin boundaries equalize the probability of the first registered detection across bins, conditioned on at least one detection within a reference cycle. Thus, our contribution is an analytical time-to-number mapping that theoretically produces uniform raw outputs without requiring customized pulse shapes or heavy attenuation. A detailed comparison is presented in Table~\ref{tab:comparison_qrng}.
\begin{table*}[htbp]
\label{tab:comparison_qrng}
\centering
\caption{Comparison of existing QRNG approaches with the present work.}
\renewcommand{\arraystretch}{1.3}
\begin{tabular}{|p{2.75cm}|p{5.6cm}|p{6.6cm}|}
\hline
\textbf{QRNG approach}&\textbf{Main Principle}&\textbf{Distinction of the present work} \\
\hline
Conventional Photon-arrival-time QRNGs~\cite{nie2014practical,wayne2009photon,ma2005random}&Random numbers are generated by measuring photon arrival times and assigning detection events to uniform time bins.&Nonuniform time-bin boundaries are analytically derived to equalize the probability of the first registered detection across bins.\\
Pulse-shaping QRNGs~\cite{wayne2010lowbias,nguyen2018programmable}&Optical pulse shapes are engineered to modify the photon-arrival-time distribution and reduce output bias.&A conventional coherent source is used, without customized pulse shaping.\\
Bias reduction and certification approaches~\cite{wahl2011ultrafast,zhang2021simple}&Techniques are developed to reduce bias to improve the reliability and certifiability of arrival-time-based random-number generation.&Closed-form bin boundaries are derived to theoretically equalize the first-detection probabilities.\\
Bounded-bias QRNGs~\cite{wahl2011ultrafast}&The output bias is characterized and constrained within a specified bound.&The present work aims at theoretically uniform raw outputs under the assumed detection model, rather than merely bounding the output bias.\\
Post-processing-free QRNGs~\cite{dynes2008high,nguyen2018programmable}&Approaches designed to generate random outputs without conventional post-processing, using specific photon-arrival-time or optical pulse-shaping techniques.&Theoretical uniformity is obtained through an analytically designed time-to-number mapping; practical imperfections may still require post-processing.\\
Low-bias QRNGs~\cite{wayne2010lowbias}&Output bias is reduced through optical pulse engineering.&The theoretical bias is reduced by optimizing time-bin widths rather than modifying the optical pulse shape.\\
\hline
\end{tabular}
\end{table*}

\end{document}